\documentclass[12pt,titlepage]{article}
\usepackage{setspace}
\usepackage{amsmath}
\usepackage{graphicx}
\usepackage{color}
\usepackage[round,numbers,sort&compress]{natbib}
\title{Quantitative transcription factor binding kinetics at the single-molecule level}

\author{
    Yufang Wang \\
        Department of Physics \\
        Princeton University, Princeton, NJ
    \and
    Ling Guo \\
        Department of Molecular Biology \\
        Princeton University, Princeton, NJ
	\and
	Ido Golding
    \thanks{Present address: Dept. Physics, Univ. of Illinois, Urbana, IL 61801} \\
        Department of Molecular Biology \\
        Princeton University, Princeton, NJ \\
    \and	
    Edward C. Cox
    \thanks{
        Corresponding author. Address:
        Department of Molecular Biology, Moffett Lab-333
        Princeton University, Princeton, NJ~08544, U.S.A.,
        Tel.:~(609)258-3856
    }\\
        Department of Molecular Biology\\
        Princeton University, Princeton, NJ
    \and
    N. P. Ong\\
        Department of Physics\\
        Princeton University, Princeton, NJ
}

\date{}

\begin{document}

\maketitle

\abstract{
We have investigated the binding interaction between the bacteriophage $\lambda$ repressor CI and its target DNA using total internal reflection fluorescence microscopy.  Large, step-wise changes in the intensity of the red fluorescent protein fused to CI were observed as it associated and dissociated from individually labeled single molecule DNA targets. The stochastic association and dissociation were characterized by Poisson statistics. Dark and bright intervals were measured for thousands of individual events. The exponential distribution of the intervals allowed direct determination of the association and dissociation rate constants, $k_\textrm{a}$ and $k_\textrm{d}$ respectively.  We resolved in detail how $k_\textrm{a}$ and $k_\textrm{d}$ varied as a function of 3 control parameters, the DNA length $L$, the CI dimer concentration, and the binding affinity. Our results show that although interaction with non-operator DNA sequences are observable, CI binding to the operator site is not dependent on the length of flanking non-operator DNA.

}

\clearpage

\section*{Introduction}
The CI protein, the product of the \emph{cI} gene of bacteriophage $\lambda$, has long served as an important model system for understanding protein-DNA interactions and gene regulation~\citep{Ptashne1980, Johnson1981}. In $\lambda$ phage, CI monomers assemble into active dimers, CI$_2$'s, that bind to three DNA sites (O$_\textrm{R}$1, O$_\textrm{R}$2 and O$_\textrm{R}$3) on the right operator O$_\textrm{R}$, which determines genetic switching between the lysogenic and lytic lifestyles. Adjacent dimers interact cooperatively, and these higher order complexes can interact with similar complexes bound to 3 related tandem operators O$_\textrm{L}$1, O$_\textrm{L}$2 and O$_\textrm{L}$3 $\sim$2.4 kilo basepairs (bp) away~\citep{Ptashne1980}. Understanding the dynamics of binding is important for unraveling the relevant physical mechanism of genetic control, as well as broader problems such as macro-molecular recognition.

The interactions between CI and its targets may be the best studied repressor-operator system in molecular biology~\citep{Reichardt1971, Meyer1975, Strahs1994, Merabet1995, Anderson2008}.  However, we know nothing about how individual stochastic binding events sum to yield the ensemble averages revealed by nitrocellulose filter binding assays~\citep{Kim1987}, DNase footprinting~\citep{Tullius1986}, electrophoretic mobility shifts~\citep{Vipond1995}, and isothermal titration calorimetry~\citep{Merabet1995}.
In recent years, several single-molecule techniques have been introduced to study protein-DNA interactions~\citep{Nie1997, Schutz1998, Xie1999, Weiss2000}. The tethered-particle motion technique~\citep{Zurla2006} provides dynamic information about single-molecule length at 100 nm scale. Fluorescence-aided single molecule sorting~\citep{Kapanidis2004, Nalefski2006} provides the fraction of molecules bound in the reaction at the single molecule level. A very promising technique is single-molecule fluorescence microscopy (SMFM)~\citep{Taguchi2001}, which is capable of imaging the same molecule for long periods. In order to study target specificity and binding on a molecule-by-molecule basis, we used the SMFM method, tethering the target DNA individually labeled with single fluorophore to a surface, and labeling the CI protein by fusion to a red fluorescent protein, tdimer RFP~\citep{Campbell2002}.

As noted above, because each DNA molecule can interact with several CI$_2$ molecules, this interaction may have multiple outcomes, with variable numbers of CI$_2$ molecules interacting with their targets and each other -- the regulatory circuit is complex. To investigate these interactions at the most basic level, we restricted our studies to the binding of CI$_2$ to a single operator sequence, O$_\textrm{R}$1, thus avoiding the complexities of these higher order interactions.

In experiments of these kinds, large uncertainties often arise from difficulties in comparing substrate and protein concentrations [CI$_2$] from sample to sample, and experiment to experiment. We addressed this problem by arranging different DNA populations in an array geometry on the same substrate and immersing the DNA in the same protein solutions. This allowed one variable -- e.g. the DNA length $L$ -- to be varied while holding [CI$_2$] fixed. With these DNA arrays, we observed reproducible variations of the dissociation and association rate constants with respect to both $L$ and [CI$_2$].

Results from DNA molecules with and without O$_\textrm{R}$1 sites, with O$_\textrm{R}$1 flanked by a range of different nonspecific DNA lengths, and with CI repressors from both wild-type and a tight-binder mutant were compared.  Large step-wise changes in fluorescence intensity enabled us to resolve clearly the association and dissociation of single CI$_2$ molecules from their targets.  This proved valuable in the stochastic analysis of the intensity vs. time traces. We measured dissociation as well as association rate constants $k_\textrm{d}$ and $k_\textrm{a}$. The values we measured agree qualitatively, although not quantitatively, with those obtained in bulk assays, which provide ensemble averages. We found that operator specific rate constants had very little dependence on DNA length. With increasing affinity of a mutant repressor for its target, the dissociation rate constant becomes smaller while the association rate constant increases. We also found that although CI binding to nonspecific DNA sequences is easily observed, nonspecific DNA flanking the O$_\textrm{R}$1 sequence has no influence on specific rate constants, suggesting that 1D searching along nonspecific sequences is not an important aspect of repressor targeting over the DNA length $L$ studied here.

\section*{Materials and Methods}
\subsection*{DNA and CI}
The O$_\textrm{R}$1 sequence was inserted into the pGEM\textsuperscript{\textregistered}-7Zf(+) vector (Promega Corporation, Madison WI) at the ApaI site. The local sequence is $5'$-...{GGG CCC} TTA TCA CCG CCA GAG GTA {GGG CCC}...-$3'$, where {GGG CCC} is the ApaI recognition site. The construct was verified by sequencing. The polymerase chain reaction (PCR) was used to amplify the O$_\textrm{R}$1 sequence together with the flanking sequences using the Expand HiFi PCR System (Roche Diagnostics Corporation, Indianapolis, IN). The PCR product was purified with Qiagen QIA PCR Kit (Qiagen Inc. Valencia, CA). As shown in Fig.~\ref{fig:Intro}A, a common left primer ($5'$-CAG TGA ATT TGT AAT ACG ACT CAC TAT AGG G-$3'$) was used with different right primers (Table~\ref{primertable}) to create DNA of varying lengths. The common left primer has a biotin group attached to the 5' end to immobilize the DNA.
A green fluorescent Alexa Fluor\textsuperscript{\textregistered} 488 was attached at the base to the ninth T from the $5'$ end of the left primer. All primers, including the modifications were ordered from Integrated DNA Technologies, Coralville, IA. In particular, the internal Alexa Fluor 488 dT modification is available through off-catalog order. We call the DNA with the O$_\textrm{R}$1 sequence operator (op) DNA. We also made DNA molecules without the operator sequence using the original vector as PCR template. We call them non-operator (nop) DNA. DNA lengths were verified using electrophoresis.

All CI protein preparations used in our single molecule investigations were fusions with tdimer (a red fluorescent protein, tdimer(2)12 in~\citep{Campbell2002}). HN-tagged WT CI constructs, Lys-34 CI, Cys-88 CI and their respective tdimer fusions were cloned into the PROTet$^{\rm TM}$ 6$\times$HN cloning vector by standard methods~\citep{Sambrook1989}. All plasmid inserts were verified by sequencing. CI function was established by demonstrating that strains carrying each construct in the PROTet$^{\rm TM}$ vector (CLONTECH Laboratories, Inc. Mountain View, CA) displayed the expected sensitivity to $\lambda vir$ and $i^{434}$, and resistance to  $\lambda$ PaPa superinfection, hallmarks of functional CI.  CI and the tdimer:CI fusions were then purified by Talon$^{\rm TM}$ column chromatography (CLONTECH Laboratories, Inc. Mountain View, CA). The molarities of purified fractions were established by quantitative densitometry of each preparation by comparison to known quantities of bovine serum albumin run in adjacent lanes of SDS-polyacrylamide gels. Molecules of tdimer:Cys-88 CI, whose threonine residue at position 88 had been replaced by a cysteine, were shown to be dimers by electrophoresis in non reducing SDS-polyacrylamide gels, as expected~\citep{Sauer1986}. Purified fractions were tested for \emph{in vitro} activity by standard gel-shift assays using a 200 bp DNA fragment containing the $\lambda$ O$_\textrm{R}$ operator consensus sequence and non-specific competitor DNA as a control for sequence specificity~\citep{Chen2005}. The tdimer:CI fusions were as active or more so than respective CI proteins on a molar basis.

The tdimer:CI preparations always contained a contaminating protein with a molecular mass of $\sim$25 kD.  Because this is close to the mass of native CI, which would act as a nonfluorescent competitor in our experimental system, we cut out both the tdimer:CI band and the $\sim$25 kD band from a denaturing polyacrylamide gel and identified them by mass spectrometry.  The fusion product exhibited the expected peptides from the tdimer and CI.  There was no CI signal in the contaminating band.  The identified peptides were entirely from the tdimer.  We suspect that this contaminant arises either by proteolysis (unlikely) or premature chain termination of the tdimer as it is synthesized.

\subsection*{Preparation of DNA arrays in channels}
DNA molecules were immobilized on coverslip surfaces. Glass surface treatments follow~\citep{Pal2005}. Coverslips (22 mm $\times$ 40 mm) were cleaned by first soaking in 5\% Contrad\textsuperscript{\textregistered} 70 (Decon Laboratories, Inc. King of Prussia, PA) for 2 hr, followed by rinsing and immersing in deionized water for 1 hr. They were then rinsed with and immersed in acetone for 5 min., followed by 5 min. in Vectabond\textsuperscript{\texttrademark} (Vektor Laboratories, Burlingame, CA) solution (2\% v/v in acetone).  They were then washed with acetone and deionized water and blow-dried with clean compressed air. The coverslip surfaces were exposed to polyethylene glycol (PEG) coating solution, which was prepared as 25\% (w/v) mPEG-SPA (Nektar Therapeutics, Huntsville, AL) and 0.25\% (w/v) biotin-PEG-NHS (Nektar Therapeutics, Huntsville, AL) in 50 mM NaHCO$_3$ (pH 8.5). After incubating in a humid chamber for 3 or more hr, the coverslip surfaces were rinsed with deionized water and kept in deionized water for an additional 2 hr. After blow drying, the coverslips were exposed to 0.1 mg/ml streptavidin solution for 15 min, rinsed with deionized water and blow-dried again.

\begin{figure}
   \begin{center}
      \includegraphics*[width=7in]{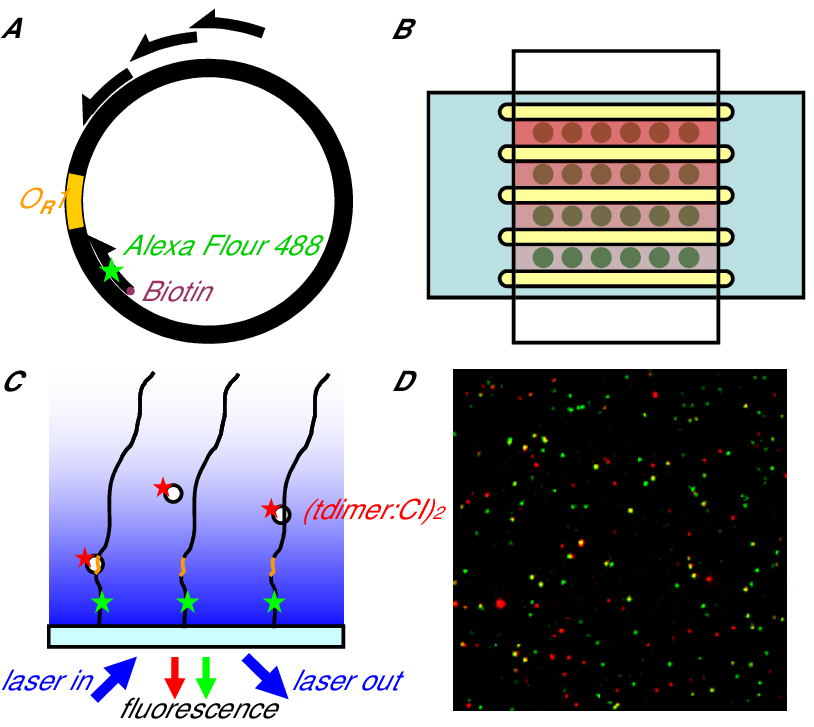}
      \caption{}
      \label{fig:Intro}
   \end{center}
\end{figure}

Solutions of biotin labeled DNA were pipetted onto a coverslip to form a typically $4\times 6$ array of patches (Fig.~\ref{fig:Intro}B). After a 15-min. incubation in a humid chamber, DNA molecules were immobilized by biotin-streptavidin interaction. To avoid mixing between different DNA populations, excess solution on the coverslip was blotted dry with bibulous paper. Binding buffer (10 mM PIPES, 50 mM KCl, 5\% DMSO, 0.1 mM EDTA) was pipetted to the same spots and blotted dry repeatedly to remove unbound DNA. We divided the array into 4 channels with walls of vacuum grease.  The channels were capped with a second coverslip coated with PEG and filled with tdimer:CI solution in binding buffer before sealed with beeswax.

Protein absorption on the glass surfaces was minimized by the PEG coating. Non-specific absorption could change the actual protein concentration at low protein concentration. We observed that protein absorption was at most one molecule per $\mu$m$^2$. The liquid layer was about 200 $\mu$m thick. At the lowest CI$_2$ concentration of 70 pM, we estimate that non-specific binding could decrease the concentration by at most 20$\%$, which would lower the measured association rate constants by only a small amount.

To minimize tdimer:CI denaturation, protein solutions filled into the channels were freshly diluted $\sim$1000-fold into binding buffer from a small amount of the $\mu$M stock solution just prior to each experiment. Concentrations of CI$_2$ were calculated assuming a monomer-dimer dissociation constant of 20 nM~\citep{Burz1994}. By following this protocol, each patch of DNA in the same channel in Fig.~\ref{fig:Intro}B was exposed to the same [CI$_2$]. From one channel to the next, [CI$_2$] increased by a factor of $\sim$3. This allowed us to reduce significantly problems associated with differences in surface preparation and protein concentration. Hence, the 24 patches with varying $L$ and [CI$_2$] were investigated on the same substrate in one experimental run. The patches of DNA molecules were incubated with CI solution for $\sim$60 min before acquiring fluorescence microscopy images.

Time-lapse series were captured for all DNA lengths immersed in a selected CI concentration for each prepared sample. A CI concentration that is too low resulted in very few binding events, while a concentration that is too high resulted in multiple CI$_2$ binding to a single DNA. The CI concentration that maximized single-CI$_2$ DNA interactions was selected. An experimental run was typically five to six hours for each prepared array. During each run six 40- or 60-min time-lapse image series were acquired. The CI protein degradation was insignificant as judged by fluorescence intensity during the experiment.

\subsection*{Single Molecule Fluorescence Microscopy}\label{SMFM}
DNA was immobilized to the coverslip surface and immersed in CI solution. Snapshots and time-lapse images were captured at room temperature by total internal reflection fluorescence (TIRF) microscopy (Fig.~\ref{fig:Intro}C) with a Nikon TE2000-U inverted microscope. Automated image acquisition was made possible with HF204 emission filter wheel and H112 focus drive (Prior Scientific Inc. Rockland, MA). The sample was illuminated with the argon 488 nm laser line incident near the critical angle.

The evanescent wave depth $\delta$ equals $\frac{\lambda}{2n\pi \sqrt{\sin\theta_i-\sin\theta_\textrm{c}}}$, where $\lambda$ is the light wavelength in vacuum, 488 nm, $n$ is the refractive index of glass, 1.52, $\theta_\textrm{c}$ is the critical angle from glass to water, 61.4$^\circ$. The excitation intensity at the glass-water interface is also a monotonically decreasing function of the incident angle that can be theoretically calculated~\cite{JacksonEM}. We calibrated the incident angle by measuring the intensities of 20 nm fluorescent spheres immobilized on a glass surface. We estimated that the evanescent wave depth was 300 nm or deeper in our experiments. Thus the entire length of the DNA was illuminated. We also note that the longer DNA's, especially those longer than the 200 bp persistence length, describe a volume that is on average closer to the surface, and hence on average the bound CI molecules were in the evanescent field.

Both Alexa Fluor 488 and tdimer were excited at 488 nm. The emission of Alexa Fluor 488 peaks around 517 nm (green), while the emission of tdimer peaks around 579 nm (red). Dichroic mirror, green and red emission filters were used to separate excitation light and fluorescence from different molecules (Z488rdc, HQ525/50m, HQ645/75m, Chroma Technology, Rochingham, VT). Images were recorded with an EM-CCD camera (DV877-DCS-BV, Andor Technology, South Windsor, CT). To minimize fluorophore bleaching, the laser power density was kept at $\sim$19 W/cm$^2$ over an illuminated area of $\sim$ 60 $\times$ 40 $\mu$m$^2$.

For still images, a green frame was acquired immediately followed by the acquisition of a red frame, each with 1 s exposure time. For time-lapse images, a single green frame was only taken once at the beginning of each time-lapse run with 1 s exposure time, since the DNA molecules were immobilized. Red time-lapse images were acquired with 0.1 s exposure time every 10 or 15 s. The sample is only illuminated while a picture is being taken to minimize tdimer bleaching and therefore to maximize event recording. In our experiment, the time it takes for a bound CI$_2$ to dissociate from DNA was on the order of minutes. We chose [CI$_2$] so that it took about the same amount of time for an unbound CI$_2$ to associate with DNA. In the supplemental movie, the DNA molecules (marked with blue circles) could be seen dynamically interacting with CI$_2$.

\subsection*{Image analysis}
First, the background fluorescence was subtracted from every image. The background fluorescence mainly comes from the buffer autofluorescence, which is proportional to the illumination intensity. The laser illumination has a two dimensional (2D) Gaussian profile. The image background was determined by fitting the image, most of whose pixels were background pixels, to a 2D Gaussian function. After background subtraction, the image is also normalized by the same 2D Gaussian profile to compensate for the nonuniform illumination. Only the center of the image, the region with the highest signal to noise ratio, was used for further analysis.

In each image, fluorescent dye or protein molecules appear as diffraction-limited bright spots. These spots were identified by thresholding. The intensity of a bright spot was calculated as the average intensity of the 3 by 3 pixels around the local maxima above background. In the still images, the red intensity was normalized by the mean intensity of tdimer:Cys-88 CI molecules, which form covalent dimers independent of concentration~\citep{Sauer1986}, as dimers are the smallest active units interacting with DNA. From now on, we will mention tdimer:CI as CI in this paper, since all our CI proteins are tdimer fusions.

Cross-correlation was calculated between the green DNA image and the red CI image. Colocalizeation of CI$_2$ and DNA resulted in strong cross-correlation peak, which was also used to align the green and red images. By overlaying the aligned green image over the red, we easily identified the binding events of CI$_2$ to DNA as the colocalization of green and red spots, which acquired a yellow hue.

Time-lapse images were also aligned to correct horizontal drift. For time-lapse images, it was often observed that at the locations of green DNA molecules, the red intensity showed step-wise switches between zero and non-zero levels. We identify these changes with the association and dissociation of red fluorescent CI$_2$ to and from DNA.

A semi-automated computer program was developed to divide the red intensity vs. time curve ($I(t)$) into a series of relative flat segments connected by sharp transitions. Before analyzing the raw trace of red intensity $I(t)$, we measured two intensities $I_0$ and $I_1$ in several manually selected traces. $I_0$ is the average quiescent intensity when no CI$_2$ proteins are bound to the DNA. $I_1$ is the average intensity when only one CI$_2$ is bound. The values of $I_0$ and $I_1$ are reproducible from run to run and are well defined throughout the experiment. Assuming the intensity is linearly scaled with number of CI$_2$ bounded, the average intensity of more than one CI$_2$ bound to the DNA can be estimated. Given the raw trace of the intensity $I(t)$ vs. $t$, the program achieves automatic segmentation in two steps. With the average intensities $I_n$ ($n=0, 1, 2$\ldots) known, the program readily identifies the portions of $I(t)$ belonging to no CI$_2$, one CI$_2$ and multiple CI$_2$ states. However, at this stage, the actual transition time between these states may only be estimated roughly.

To fix the transition time accurately, we employ a minimum-fluctuation method. At such low fluorescence level of a few CI$_2$ molecules, the recorded intensity fluctuation is dominated by the dark current of camera. The intensity root-mean-square (rms) amplitude $\sigma_0$ is constant regardless of the CI$_2$ number, as long as the CI$_2$ number remains small and constant. In a raw trace $I(t)$, when a transition occurs (CI$_2$ number changes), the rms amplitude increases significantly. We consider $I(t)$ in a time interval $[t_1, t_2]$. If the rms amplitude $\sigma$ within $[t_1, t_2]$ is less than $2\sigma_0$, this interval is treated as a continuation of the previous state. However, if $\sigma \geq 2\sigma_0$, the program recognizes that a transition has occurred. It then picks an arbitrary point $t_3$ within $[t_1, t_2]$, and calculates the rms amplitude $\sigma_1$ within $[t_1, t_3]$ and $\sigma_2$ within $[t_3, t_2]$. By minimizing the quantity $\frac{(t_3-t_1)\sigma_1^2+(t_2-t_3)\sigma_2^2}{t_2-t_1}$ with respect to $t_3$, the program then determines accurately the actual transition time. In addition we also recognize small clusters of $I(t)$ data points with significantly different mean values from the surrounding data points as very short segments.

Because low signal to noise level can confuse the program, the segmented curves were visually checked and manually corrected if necessary. Only segments that begin and end within the complete trace of $I(t)$ were used for our histograms.

\subsection*{Gel-shift assay}
The same fluorescent DNA and CI in single-molecule experiments were used for the gel-shift assays. DNA and CI were mixed in binding buffer supplemented with 100 $\mu$g/ml bovine serum albumin (BSA). The binding buffer is also the buffer used in the single-molecule measurements. The concentration of the DNA was 20 nM in every reaction while the concentrations of CI$_2$ varied (either 5, 36.5 or 452 nM). The reaction mixtures were incubated at room-temperature. After about an hour, 20 $\mu$l reaction mixtures were electrophoresed in Bio-Rad 4-20$\%$ Gradient TBE gels (Bio-Rad, Hercules, CA) in a 4$^\circ$C cold room. The gels were stained with ethidium bromide for DNA. The gel images were taken under UV illumination.

\subsection*{Stochastic series}
To analyze how the intensity time trace $I(t)$ is informed by the transition rates and to define some terms, we describe transitions between states $DP_n$ (defined as a DNA bound by $n$ CI$_2$'s) as the time-dependent stochastic series
\begin{equation}
DP_0\mathop{\rightleftharpoons}_{\mu_1}^{\lambda_0}DP_1\mathop{\rightleftharpoons}_{\mu_2}^{\lambda_1}DP_2
\cdots\mathop{\rightleftharpoons}_{\mu_n}^{\lambda_{n-1}}
DP_{n} \mathop{\rightleftharpoons}_{\mu_{n+1}}^{\lambda_{n}}\cdots
\label{eq:DP}
\end{equation}
where $\lambda_n$ and $\mu_n$ are the forward and backward transition rates governing the time-evolution of the probability of finding a complex $DP_n$ at time $t$.

\begin{figure}
   \begin{center}
      \includegraphics*[width=7in]{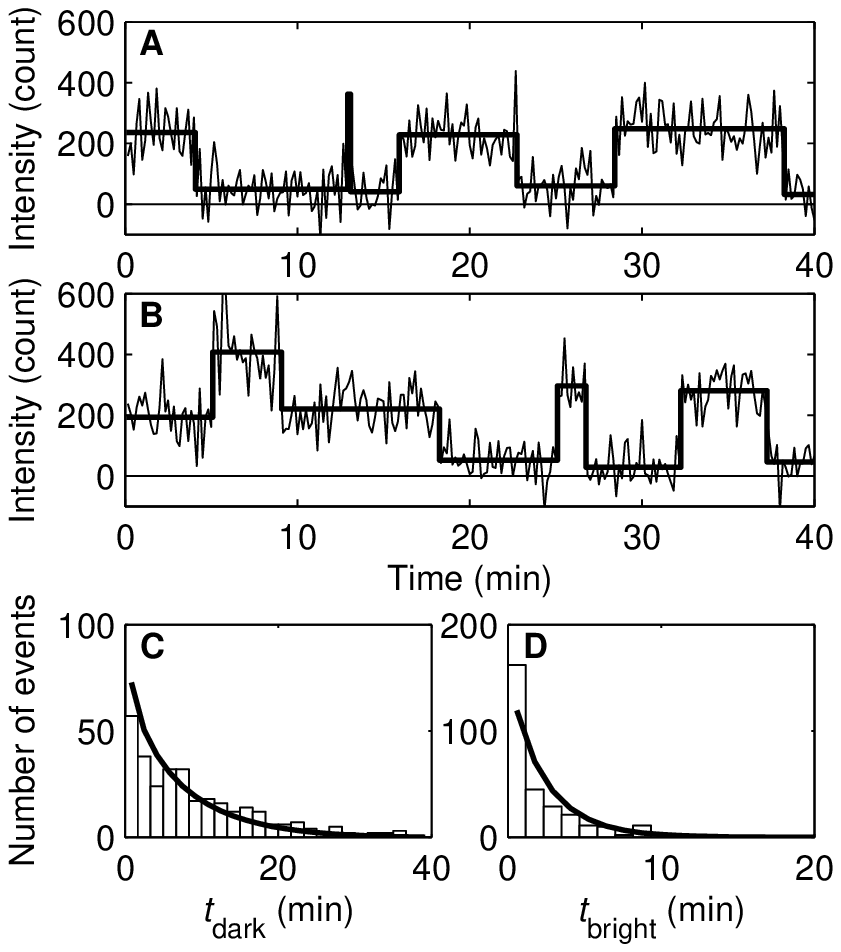}
      \caption{}
      \label{fig:trace}
   \end{center}
\end{figure}

A powerful feature of dynamic SMFM is that it allows a single transition in the series Eq.~\ref{eq:DP} to be studied in isolation. In general, the time trace $I(t)$ at a colocalized site shows both single- (Fig.~\ref{fig:trace}A) and multiple-step transitions (Fig.~\ref{fig:trace}B) relative to the background intensity. We used a semi-automated computer program that efficiently distinguishes large, step-wise transitions from small-amplitude fluctuations.

By collecting the set of traces exhibiting single-step transitions, we accumulated statistics on the transition $DP_0 \rightleftharpoons DP_1$. The ``up'' and ``down'' transitions separates the bright durations $t_\textrm{bright}$ and dark durations $t_\textrm{dark}$, whose probability distribution can be described by ${\cal P}(t_\textrm{dark}) = \lambda_0 e^{-\lambda_0 t_\textrm{dark}}$ and ${\cal P}(t_\textrm{bright}) = \mu_1 e^{-(\mu_1+\lambda_1) t_\textrm{dark}}$, respectively (Fig.~\ref{fig:trace}C,D). From the state $DP_1$, the molecule complex goes to $DP_0$ with rate constant $\mu_1$. $DP_1$ also goes to $DP_2$ with rate constant $\lambda_1$. During the same period of time, the number of $DP_1$ to $DP_0$ transitions was counted as $N_{10}$ while the number of $DP_1$ to $DP_2$ transitions was counted as $N_{12}$. We then have $\frac{N_{10}}{N_{12}}=\frac{\mu_1}{\lambda_1}$. From the individual traces of $I(t)$ at each DNA bound to CI$_2$ (200 $\sim$ 300 in the field of view), we accumulated hundreds of measurements of $t_\textrm{dark}$ and $t_\textrm{bright}$ in a single time-lapse data set. Assuming Poisson processes, $\lambda_0$ and $\mu_1+\lambda_1$ was obtained by fitting the distributions of $t_\textrm{dark}$ and $t_\textrm{bright}$ to exponential distributions according to the unbiased estimation in~\citep{Koster2006}. We obtain $\mu_1$ by multiplying the second exponential constant with $\frac{N_{10}}{N_{10}+N_{12}}$. Hence the histograms of $t_\textrm{dark}$ and $t_\textrm{bright}$ in the filtered set allow $\lambda_0$ and $\mu_1$ to be measured directly (Fig.~\ref{fig:trace}C,D).

If the transitions to $DP_2, DP_3, \cdots$ are negligible, the association and dissociation rate constant $k_\textrm{a}$ and $k_\textrm{d}$, of the reaction
\begin{equation}
DP_0 + \textrm{CI}_2 \mathop{\rightleftharpoons}_{k_\textrm{d}}^{k_\textrm{a}} DP_1
\end{equation}
are just $k_\textrm{a} = \lambda_0 /[\mathrm{CI}_2]$, $k_\textrm{d} = \mu_1$, with the dissociation constant $K_\textrm{D} = k_\textrm{d}/k_\textrm{a}$.

\section*{Results}

\subsection*{Snapshots and time-lapse images of CI interacting with DNA}
In a typical snapshot with the red channel image overlayed on the green channel image, there are many bright spots (Fig.~\ref{fig:Intro}D). The color of bright spots fall into three easily separable classes: green, red and yellow, identified as DNA, CI and DNA-CI, respectively.

The TIRF technique allows us to capture simultaneously the transitions in each of the DNA molecules within the field of view. When observing single fluorophore-labeled molecules, the blinking and bleaching of fluorophores might interfere with the measurements.

\begin{figure}
   \begin{center}
      \includegraphics*[width=7in]{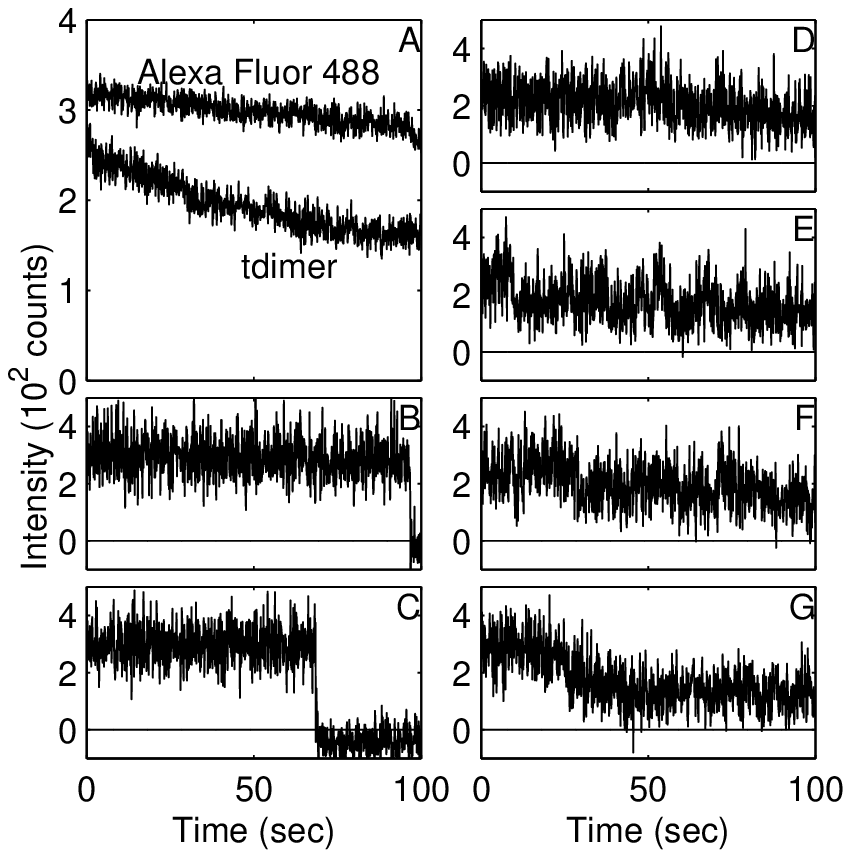}
      \caption{}
      \label{fig:bleach}
   \end{center}
\end{figure}

Molecule bleaching was moderate in our experiments. The average intensity of all fluorescent molecules in the field of view decreased by 10 to 30$\%$ after 100 seconds of continuous illumination (Fig.~\ref{fig:bleach}A). The intensities of single Alexa Fluor 488 molecules were very stable, except for the occasional characteristic single-step bleaching events (Fig.~\ref{fig:bleach}B, C). The fluorescent intensity of tdimer is more critical. Blinking and bleaching of tdimer might be confused with binding and unbinding of CI$_2$ from DNA. To control for this possible problem, tdimer:Cys-88 CI molecules non-specifically bound to a glass surface were investigated. We could identify single tdimer:Cys-88 CI dimer molecules as individual bright spots with similar intensities. With time, the intensities of these spots decreased gradually (Fig.~\ref{fig:bleach}D to G). It is worth emphasizing that the molecules were only illuminated when the image was acquired during the process of time-lapse image acquisition. The accumulated exposure time during a 40- or 60-min movie was about 100 s. With tdimer bleaching slowly and gradually during this period of time, the binding and unbinding of CI$_2$ can be easily resolved.

In the dynamic experiments, we captured 40 or 60 min movies at each of the 6 patches in the same channel.  Fig.~\ref{fig:trace}A shows a typical time trace of the fluorescence signal from one molecule over a 40 min interval.  The baseline (minimum signal intensity) corresponds to the state with no CI$_2$ bound to DNA, while the plateaus at finite intensity represent the state with one CI$_2$ bound to DNA. At very low [CI$_2$], most DNA has not bound CI$_2$ over the duration of data acquisition. However, at high [CI$_2$], several finite intensity states were observed in a single time-trace, resulting from multiple CI$_2$ interacting with a single DNA molecule. We adjusted the [CI$_2$] to optimize the number of single CI$_2$ binding events. In each field of view, there are time traces for $\sim$200 molecules on average, each displaying 1 -- 10 bright states.  As explained above, the ``filtered set'' only contains the traces with single-step transitions. The histograms of the sets of $t_\textrm{dark}$ and $t_\textrm{bright}$ in Fig.~\ref{fig:trace}C and D are well fitted by an exponential distribution, consistent with a Poisson processes and implying that association and dissociation are best modeled as single step processes. This then allows $\lambda_0$ and $\mu_1$ in each of the populations to be determined.

In some data sets, similar dark-bright red intensity switching patterns were observed at locations where there was no green fluorescence. Some of these locations might have harbored a bleached DNA molecule, while others might simply have been CI molecules non-specifically bound to the surface. To make sure we were looking at the interaction of CI$_2$ to DNA, we counted events only at locations where green fluorescence was high.

\subsection*{CI binds DNA both with and without the operator sequence, showing different binding profiles}
We first discuss measurements of binding profiles for single DNA molecules by looking at still images.

\begin{figure}
   \begin{center}
      \includegraphics*[width=7in]{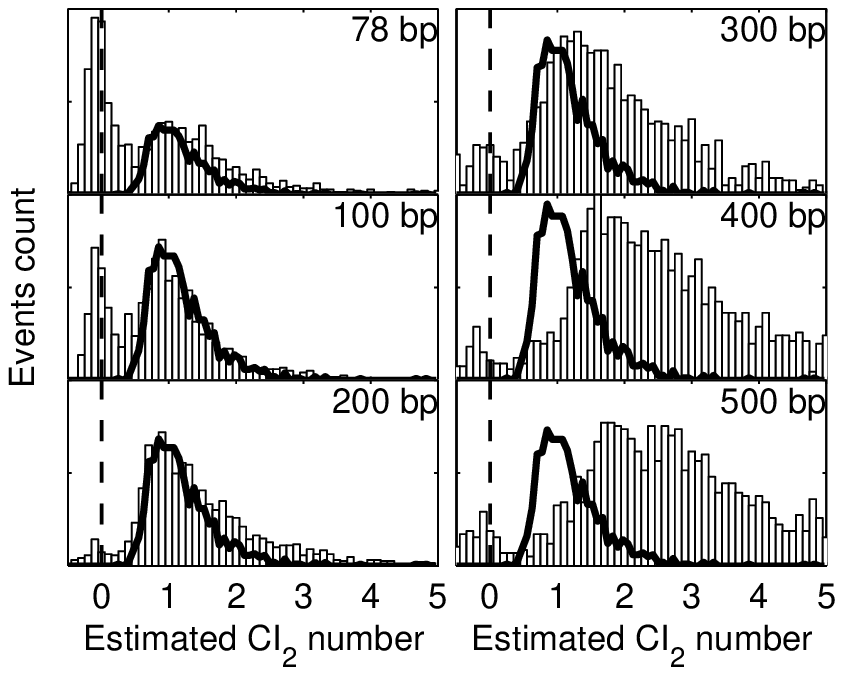}
      \caption{}
      \label{fig:hist}
   \end{center}
\end{figure}

Recall that the red intensities have been normalized using the intensity of single CI$_2$ molecules. A series of intensity histograms is shown in Fig.~\ref{fig:hist} for CI$_2$ interacting with op DNA. The histogram consists of two peaks, a narrow one centered around the background intensity (dashed line) identifies the population of DNA without CI$_2$ bound to it. For DNA from 78 to 200 bp, the position of the second peak is consistent with one CI$_2$ bound to DNA. These peaks overlap with the histogram of single Cys-88 CI dimer molecules, confirming that CI bind to DNA as a dimer. With further increase in $L$, the peak broadens and shifts, implying that two or more CI$_2$'s bind to each DNA molecule as the CI$_2$ concentration increases

\begin{figure}
   \begin{center}
      \includegraphics*[width=7in]{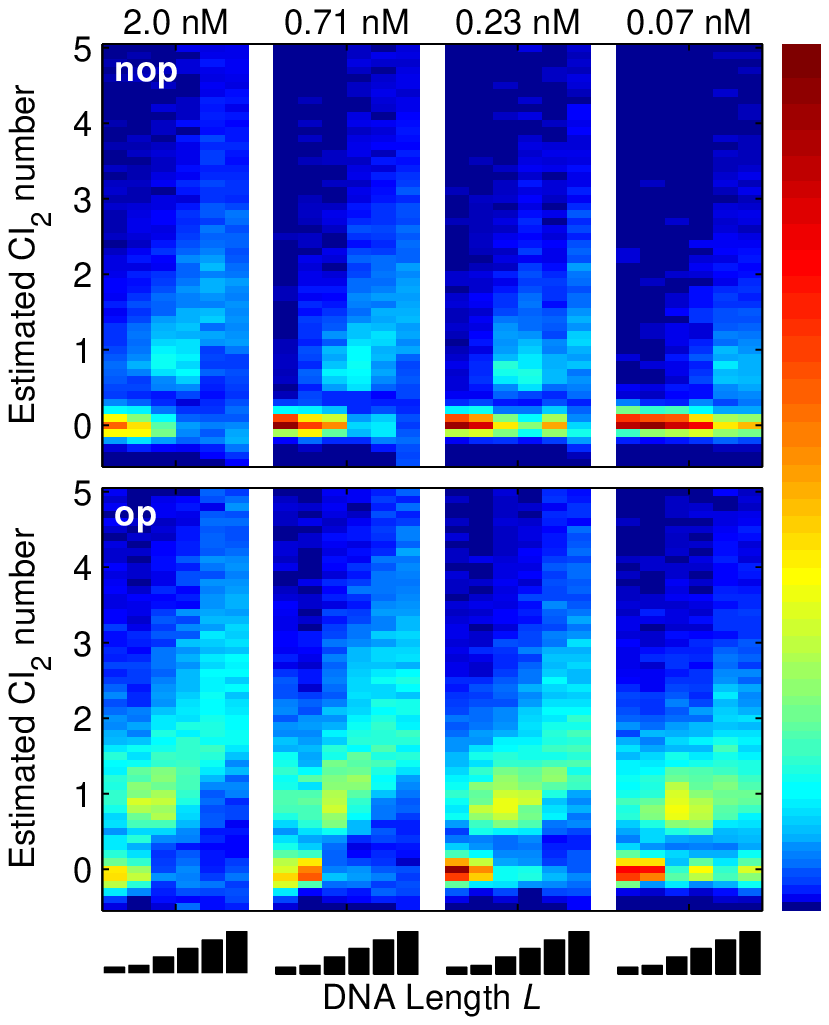}
      \caption{}
      \label{fig:L34}
   \end{center}
\end{figure}

As mentioned above, the array design allows the binding probability to be investigated as both $L$ and [CI$_2$] are varied.  A revealing way to view the trends is to display in the ($L,I$) plane the intensity histogram as a heat map, with blue indicating zero counts and red indicating large numbers of counts (Fig.~\ref{fig:L34}). The 6 vertical strips within each panel display the normalized event-count distribution in the 6 patches (with successive increasing $L$) all exposed to the same [CI$_2$]. The 4 upper panels are for nop DNA while the lower panels are for op DNA. With [CI$_2$] held fixed, more CI$_2$ was observed to bind to op DNA (lower panels), as expected from the increased affinity. Clearly, in each panel, the high intensity peak moves to larger values as $L$ increases from 78 to 500 bp.  However, in very dilute [CI$_2$] (right-most panels), short (78 $\sim$ 300 bp) nop DNA molecules mostly had no CI$_2$ to bound them, while large fraction of short op DNA had one CI$_2$ bound. It was very likely that this CI$_2$ molecule bound to DNA at the O$_\textrm{R}$1 sequence.

\begin{figure}
   \begin{center}
      \includegraphics*[width=7in]{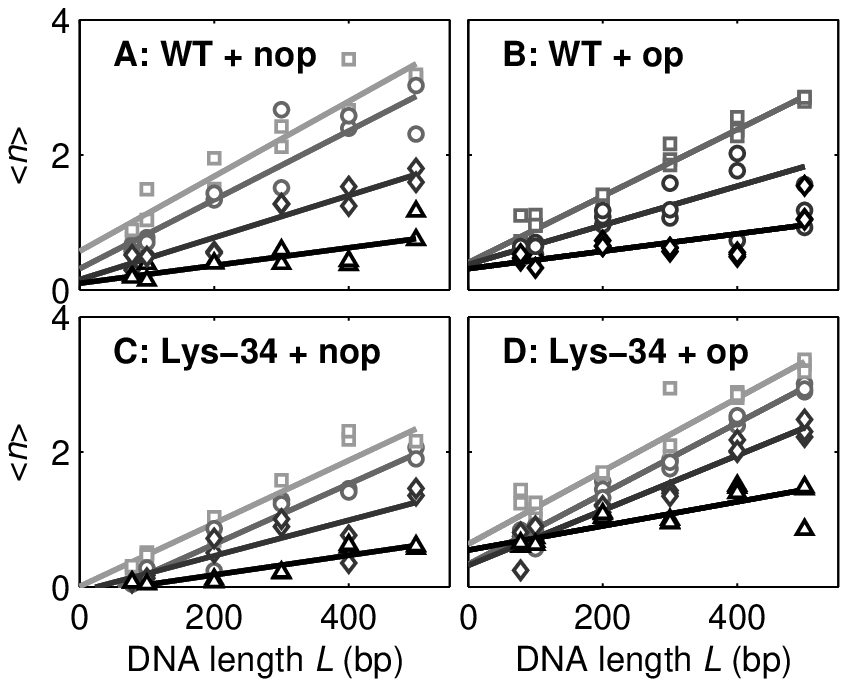}
      \caption{}
      \label{fig:mean}
   \end{center}
\end{figure}

To quantify these trends, we calculated the mean number $\langle n\rangle$ of bound CI$_2$'s per DNA molecule in each frame as the mean red intensity normalized by the intensity of single CI$_2$ molecules. Figure~\ref{fig:mean} summarizes how $\langle n\rangle$ varies with $L$ and concentration in the following 4 cases: WT CI binding to nop (Fig.~\ref{fig:mean}A) and op DNA (Fig.~\ref{fig:mean}B), and the tight-binder Lys-34 CI mutant binding to nop (Fig.~\ref{fig:mean}C) and op DNA (Fig.~\ref{fig:mean}D). In each panel, $\langle n\rangle$ is plotted against $L$ for each of the three or four [CI$_2$] (nominal values stated in Fig.~\ref{fig:mean}). The measurements were repeated 2-3 times to check reproducibility.  For a given [CI$_2$], we found that $\langle n\rangle$ increased linearly with $L$, and may be fit by a straight line. With increasing of [CI$_2$], the slope $s$ of the linear fits increased, initially in proportion to, but tending towards saturation at high [CI$_2$] (Fig.~\ref{fig:mean}A, C and D).

In general, the $L$-linear increase in $\langle n\rangle$ reflects increased availability of non-operator sites for binding. Significantly, if we look closely at the runs with op DNA (Fig.~\ref{fig:mean}B and D), we find that $\langle n\rangle$ displays a finite intercept $n_0$, viz. $\langle n\rangle = n_0 + sL$.  At low [CI$_2$], the strong dominance of binding to the O$_\textrm{R}$1 site over non-operator sites renders the association rate nearly insensitive to $L$.

\subsection*{Dissociation and association rate constants}
\begin{figure}
   \begin{center}
      \includegraphics*[width=7in]{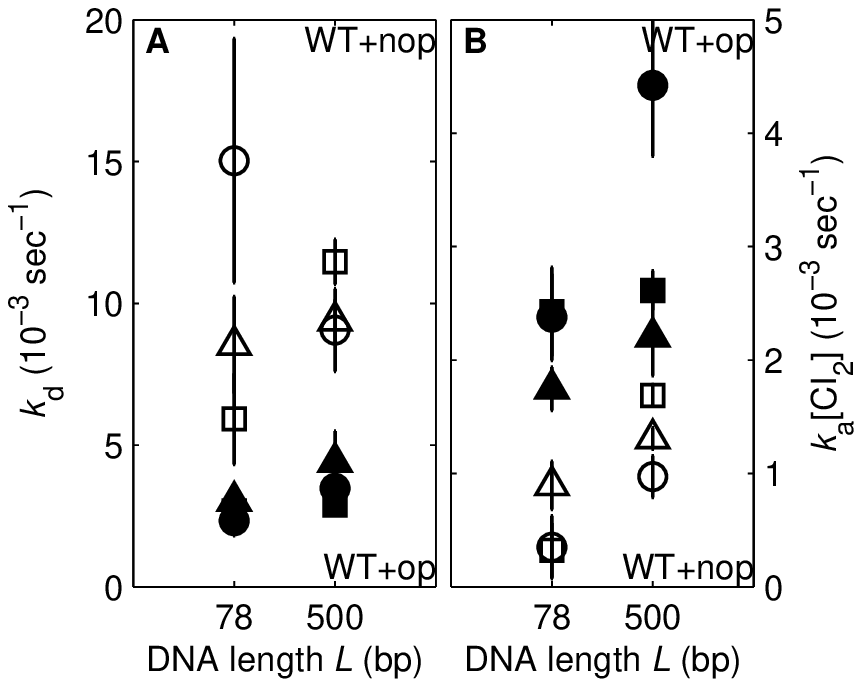}
      \caption{}
      \label{fig:comp}
   \end{center}
\end{figure}

We first compare the binding strengths of WT CI$_2$ to op and nop DNA embedded in two length extremes, 78 and 500 bp. Fig.~\ref{fig:comp}A shows the measured dissociation rate constant $k_\textrm{d}$. Values of $k_\textrm{d}$ for nop DNA (open symbols) were found to be larger than the op case (solid symbols) by a factor 2 - 4.  As expected, WT CI$_2$ binds more strongly to op DNA compared to nop. Within the uncertainty of our experiments, $k_\textrm{d}$ was not significantly different between the 2 lengths, \emph{i.e.} a 6-fold increase in $L$ had no observable effect on how long the CI$_2$ remained bound.  Three runs with nominally similar [CI$_2$] are shown.

The corresponding association rate constant $\lambda_0 = k_\textrm{a}$[CI$_2$] showed the opposite relationship between op and nop DNA. The values for $\lambda_0$ in op DNA (solid symbols) were larger than nop (open symbols) by a factor of 2 - 4 (Fig.~\ref{fig:comp}B), consistent with the higher affinity for op DNA. Moreover, the measured $\lambda_0$ for op DNA showed no significant dependence on $L$.  This is expected because binding of CI$_2$ to the single O$_\textrm{R}$1 site in op DNA dominates the op binding processes, and is independent of $L$.  By contrast, $\lambda_0$ should increase with $L$ for nop DNA. We indeed observed that $\lambda_0$ in WT+nop (open symbols) increased roughly 4-fold between the 78-bp and 500-bp populations.  We revisit this point later.

\begin{figure}
   \begin{center}
      \includegraphics*[width=7in]{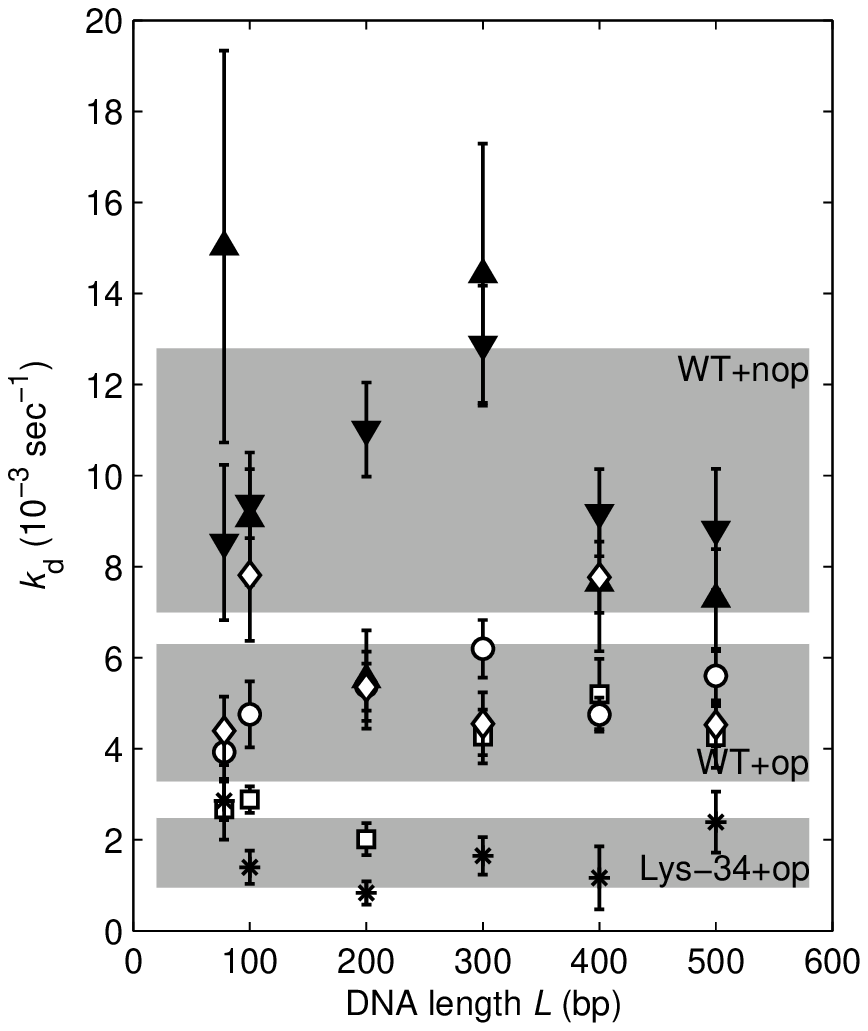}
      \caption{}
      \label{fig:kd}
   \end{center}
\end{figure}

To bolster these conclusions, we extended measurements to several DNA populations of intermediate $L$ (100, 200, 300 and 400 bp's).  In addition, we investigated the Lys-34 CI mutant, which binds more tightly than WT CI~\citep{Nelson1985}. Fig.~\ref{fig:kd} summarizes the dissociation rates $k_\textrm{d}$ measured with WT CI + nop DNA (solid symbols), WT CI + op DNA (open symbols) and for the tight-binder Lys-34 CI + op DNA (asterisks). For each DNA construct, op or nop, 6 patches of different lengths were investigated. The measured $k_\textrm{d}$'s were well separated in the 3 populations. In each case, the grey band is centered at the mean value of $k_\textrm{d}$, while its half-width gives the standard deviation. As expected, Lys-34 CI + op DNA showed the strongest binding (smallest $k_\textrm{d}$), WT CI + op DNA was intermediate, while WT CI + nop DNA had the weakest binding ($\sim$10 times weaker than Lys-34 CI + op DNA). In addition, the results strongly reinforce the conclusion that the $k_\textrm{d}$ in each case is unchanged by a 6-fold increase in $L$.

The experiments for the WT CI + nop DNA and WT CI + op DNA were run 2 and 3 times, respectively, with different sample preparations. The runs are shown as different symbols. Within the error limits, the values of $k_\textrm{d}$ were insensitive to changes in [CI$_2$], consistent with the expectation that dissociation should not be influenced by [CI$_2$].  This insensitivity to [CI$_2$] allows the data for $k_\textrm{d}$ from different runs to be meaningfully compared, as in Fig.~\ref{fig:kd}. However, $\lambda_0$, which we discuss next, is obviously $L$ dependent.

\begin{figure}
   \begin{center}
      \includegraphics*[width=7in]{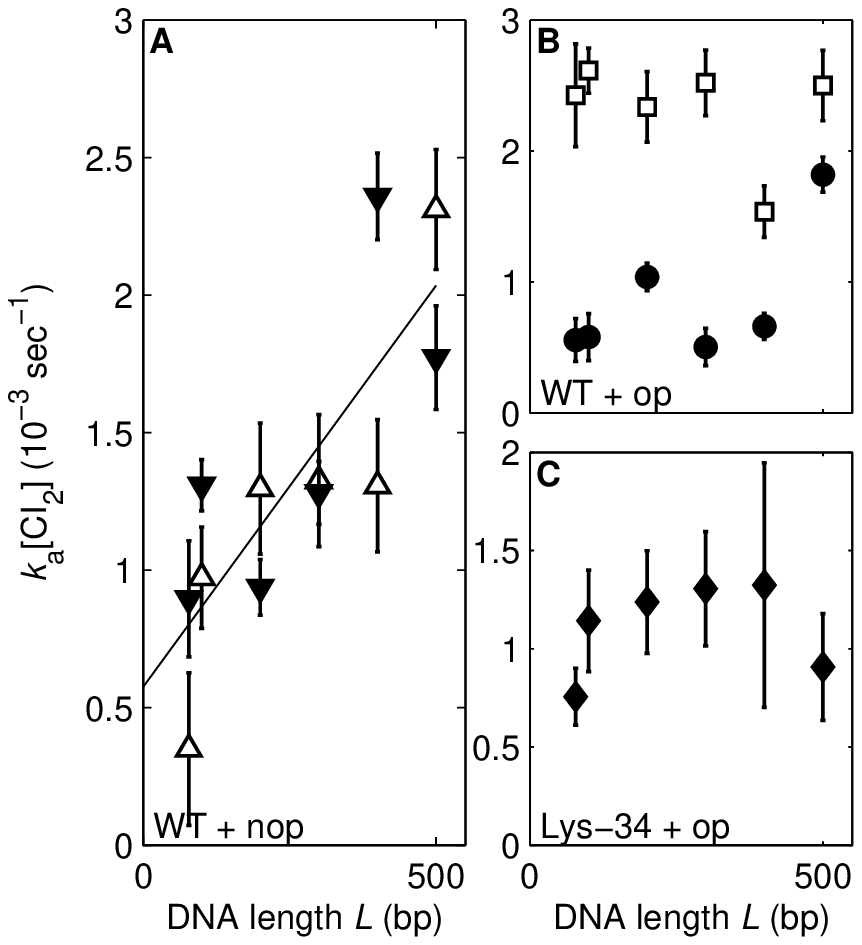}
      \caption{}
      \label{fig:ka}
   \end{center}
\end{figure}

Fig.~\ref{fig:ka} plots $k_\textrm{a}$[CI$_2$] = $\lambda_0$ obtained in these experiments. For nop DNA (Panel A), $\lambda_0$ displayed a clear increase with $L$, which confirms the trend hinted at in Fig.~\ref{fig:comp}B, and is consistent with the results of~\citep{Wang2005}.  The nominally $L$-linear increase is consistent with the expectation that a longer DNA target has a larger probability of capturing CI$_2$.  However, this is valid only for nop DNA. With op DNA, $\lambda_0$ showed no resolvable $L$ dependence for both WT CI (Fig.~\ref{fig:ka}B) and Lys-34 CI (Fig.~\ref{fig:ka}C). Apparently, the O$_\textrm{R}$1 binding site, when present, dominates any advantage that increasing $L$ might confer on repressor capture at the operator site.  In contrast with $k_\textrm{d}$, $\lambda_0$ increases by a factor of 4-5 for a $\sim$6-fold increase in [CI$_2$] (Fig.~\ref{fig:ka}) as we expect higher concentration leads to more binding events.

\subsection*{Control gel-shift assays}
The association rates measured using the single-molecule microscopy method are two orders of magnitude smaller than the values in the literature~\citep{Nelson1985}. To test if this discrepancy is caused by the experimental method or the experimental materials, gel-shift assays were performed using the same DNA and CI as well as similar buffer in the single-molecule experiments. The reaction of 20 nM DNA and three CI$_2$ concentrations, 5 nM, 36.5 nM and 452 nM, were investigated. The results are shown in Fig.~\ref{fig:gel}.

The nop DNA captured CI$_2$, which was also observed in the single-molecule measurements. About half of the DNA molecules were bound to CI$_2$ at the CI$_2$ concentration of 452 nM. The equilibrium constant $K_\textrm{D}$ for nop and CI$_2$ is estimated to be about hundreds of nM, similar to the single-molecule measurement results ($150\pm50$ nM for 78 bp nop DNA, $64\pm9$ nM for 500 bp nop DNA).

For op DNA and WT CI, about half of the DNA molecules were bound to CI$_2$ at a lower CI$_2$ concentration, 36.5 nM, in agreement with our finding that op DNA has a higher affinity for CI$_2$. The equilibrium constant $K_\textrm{D}$ is estimated from the gel-shift assays result to be about tens of nM, similar to our single molecule results of $22\pm3$ nM.

Compared with WT CI, the Lys-34 mutant of CI has higher affinity for op DNA. When CI$_2$ concentration is larger than that of DNA, almost all the DNA molecules were bound. Only at the lowest CI$_2$ concentration, 5 nM, did we observe some free DNA. We estimate that the dissociation constant $K_\textrm{D}$ to be much smaller than 20 nM. Our single-molecule measurement suggest that it is $0.11\pm0.02$ nM.

\section*{Discussion and conclusions}

\subsection*{Blinking and bleaching of the tdimer}
To investigate the behavior of tdimer:CI, we immobilized Cys-88 CI on glass surfaces without a PEG coating by non-specific binding. Recall that Cys-88 CI forms covalent dimers, which is the same as the smallest active units that binds DNA at the molarities explored here. When the protein concentration was low enough, we observed distinct bright spots with similar intensities in the red channel. With time, the intensities of these spots decrease gradually.

It is known that green fluorescent protein exhibits ``on-off'' blinking behavior on the time scale of seconds~\citep{Dickson1997}. For GFP blinking, the fractional \emph{on} time is defined as
$F_\textrm{on}=\langle\Delta t_\textrm{on}\rangle/(\langle\Delta t_\textrm{on}\rangle+\langle\Delta t_\textrm{off}\rangle)$,
where $\Delta t_\textrm{on}$ ($\Delta t_\textrm{off}$) is the intervals when the fluorescence is on(off). $F_\textrm{on}$ value depends on excitation intensity $I_\textrm{ex}$ such that
$F_\textrm{on} = I_\textrm{s}/(I_\textrm{s}+I_\textrm{ex})$.
The saturation excitation intensity $I_\textrm{s}$ equals 1.5 kW/cm$^2$ for S65T-GFP~\citep{Garcia-Parajo2000}. In our experiments, the excitation intensity is as low as 19 W/cm$^2$. If the monomers of tdimer have similar $I_\textrm{s}$, $F_\textrm{on}$ would be nearly 1, which means that the monomers would be in the on state most of the time. If the mean off time $\langle\Delta t_\textrm{off}\rangle$ is on the order of seconds, like S65T-GFP, the mean on time $\langle\Delta t_\textrm{on}\rangle$ would be as long as the total exposure time in our experiment runs.

Further studies with the red fluorescent protein DsRed showed that this closely packed tetramer protein has four distinctive intensity levels for each single molecule. Each intensity level has a life time of seconds~\citep{Garcia-Parajo2001}. However, little is known about the tdimer, which is a tandem dimer of two RFP monomers. Each CI$_2$ has two independent tdimers. One would therefore expect that (tdimer:CI)$_2$ would bleach in at lease two steps and possibly up to four steps. However, the intensity change in each step would be small comparing with photon number fluctuations in our experiments. In addition, the DNA-CI$_2$ complex has considerable rotation freedom in solution, which causes intensity fluctuation with linear polarized excitation illumination used here. For these reasons, then, we believe that the intensity loss of the (tdimer:CI)$_2$ fusion due to bleaching appears as a smooth gradual process. Hence it is reasonable to identify the rapid and step-wise changes in intensity with the unbinding of CI$_2$ from DNA.

\subsection*{Single molecule vs. ensemble measurements}
Several bulk assays have been used to investigate DNA-protein interactions. The nitrocellulose filter binding assay~\citep{Kim1987} is based on the fact that protein binds to nitrocellulose, trapping any DNA bound to it. In the DNase footprinting assay~\citep{Tullius1986} proteins protect DNA sequences from enzymatic cleavage when the proteins are specifically bound to the sequences. In gel-shift assays~\citep{Vipond1995} protein-bound DNA migrates more slowly during electrophoresis. Calorimetry~\citep{Merabet1995} is based on heat generation/absorption when proteins bind to DNA. Surface plasmon resonance (SPR)~\citep{Smith2003} has also been used to investigate the interaction between molecules.

Some of these bulk assays can provide binding profiles, while others provide binding dynamics. All but the calorimetry experiments are nonequilibrium measurements. Compared with bulk assays, an advantage of SMFM is that it can provide both the binding profile and the binding dynamics in great detail. When analyzing binding dynamics, we can filter out multiple molecule interaction, leaving only the one-on-one DNA-protein interactions. We can also study association and dissociation in isolation with the reaction system in equilibrium.

\subsection*{The observed rate constants vs. values in the literature}

The association rate constant we measured are considerably smaller than those measured with bulk assays~\citep{Nelson1985, Johnson1980}. Could this be because of our relatively slow sampling rate, since the events that occur faster than the time interval between two consecutive frames would not be detected? First, we note that most of the binding events last for many sampling frames. It is not easy to see how we could miss fast on and off rates \emph{at the same site}. Secondly, we were careful to control for this possibility by increasing our sampling rate in preliminary experiments, where we did not observe any significant changes in the observed rate constants. Increasing sampling rate caused the tdimer to bleach faster, resulting only in fewer events being recorded.

A second possibility for the discrepancy between our rates and values in the literature may be that the fusion of the bulky tdimer prevents CI from dimerization or binding to DNA. CI molecules bind to DNA via the N terminus, and dimerize through the C terminus. Without the C terminus, the N-terminal domain alone is known to have much lower affinity ($K_\textrm{D}\sim$ 1 $\mu$M) than the intact CI protein~\citep{Sauer1979}. We reason that, because the tdimer was fused to the N terminus, it should not affect CI dimerization. Moreover, according to the crystal structure~\citep{Stayrook2008}, the N-terminus points away from DNA, so the tdimer should not affect the DNA binding either. As noted in Materials and Methods, these constructs confer immunity to superinfection. They have also been used to study oligomerization by fluorescence correlation spectroscopy (FCS)~\citep{Samiee2005}. We compared the intensity of CI bound to DNA with the intensity of the Cys-88 covalent CI dimer. The Cys-88 mutant CI is confirmed to form dimer by comparing the molecular mass in both reducing and non-reducing SDS gels. As shown in Fig.~\ref{fig:hist}, the Cys-88 covalent CI dimer and tdimer:CI bound on short DNA molecules have similar fluorescence levels. In summary, the evidence suggests that these fusion products function as dimers, both \emph{in vivo} and \emph{in vitro}, and at concentrations in the nM range.

Yet another possibility is that the glass surface may somehow severely restrict accessibility of CI to the DNA molecules, despite the long distance (40 bp, $\sim$ 14 nm) between the operator sequence and the immobilization point.

To address these questions, we conducted control gel-shift assays with the same DNA and CI molecules reacting in buffer solutions similar to that used in the single-molecule experiment. The results show that the affinity of the tdimer:CI is not as high as reported in the literature (\emph{e.g.} $K_\textrm{D} \sim (2.9 \pm 0.9)\times10^{-13}$ M for op DNA and WT CI$_2$ at 50 mM KCl)~\citep{Johnson1980}, but rather similar to our single-molecule results ($K_\textrm{D} = (2.2 \pm 0.3)\times10^{-8}$ M for op DNA and WT CI$_2$). However, the affinity difference between CI binding to op and nop DNA, as well as between WT and the tight binding Lys-34 mutant was confirmed in these fusion proteins.

\subsection*{Binding dynamics suggested very short 1D search lengths}
It is widely believed that transcription factors find their targets by first associating at non-operator sequence, and then diffusing along DNA using sliding and hopping to find their targets. Since hops are assumed to cover very short distances, they can be combined with sliding by scaling up the sliding length to the length of a 1D search $s_\textrm{e}$~\citep{Kolesov2007}. The interaction of CI$_2$ with DNA can then be summarized with a 1D-3D model: CI$_2$ spend its time either by diffusing 3D in the space or by sliding 1D along the DNA. When CI$_2$ jumps on(off) the DNA molecule we will see abrupt red intensity increase(decrease) at the location of DNA. The dynamics should allow effective 1D searching lengths $s_\textrm{e}$ of $10^3$ bp to be distinguished from those $\sim$20 bp by comparing 78 to 500 bp DNA molecules.

We consider first $k_\textrm{a}$ for nop DNA. We expect it to scale with the length of DNA since doubling the DNA lengths doubles the cross section for capturing a CI$_2$. A linear fit to our data shows that $k_\textrm{a}$ increases linearly with $L$, but with an intercept $k_\textrm{a} \sim L+200$ bp. The reason for a non-zero intercept could be that immobilizing the DNA on a surface makes the two ends of DNA nonequivalent for CI$_2$ capture. If so, the positive value suggests that the immobilized end captures CI$_2$ more often than the free end.

For op DNA, one would expect to measure a length-dependent $k_\textrm{a}$ if CI$_2$ associates with any sequence and then finds its target through 1D diffusion. Instead, we observed that $k_\textrm{a}$ in op DNA exhibits little length dependence. This implies that CI$_2$ associates with the operator sequence with much higher probability than with a non-operator sequence, which is expected. It also implies that the effective 1D search length is shorter than 78 bp, the shortest DNA length in our experiments. Even so, one might still expect some length dependence since the CI$_2$'s first interaction with DNA might be non-specific. However, the operator sequence is closer to the immobilized end and is always at the same distance from the immobilization point. The result with nop DNA suggests that the immobilized end and the free end behave differently.  Because the immobilized end captures more CI$_2$, we expect to see more CI$_2$ binding to the operator sequence than CI$_2$ binding with non-specific sequence farther from the immobilizing point. Thus the length dependence of $k_\textrm{a}$ associated with non-specific interaction may be weakened.

As for $k_\textrm{d}$, we observed clearly that CI$_2$ bound to op DNA dissociates more slowly than CI$_2$ bound to nop DNA. This confirms that the operator sequence has a higher affinity. Secondly, we see that Lys-34 CI has a smaller $k_\textrm{d}$ than WT CI when binding to op DNA. This confirms that the interaction is protein specific. As shown in Fig.~\ref{fig:kd}, when affinity increases, $k_\textrm{d}$ decreases ($1 \times 10^{-2}$ s$^{-1}$ for WT CI + nop DNA compared with $5 \times 10^{-3}$ s$^{-1}$ for WT CI + op DNA), by a factor of 2. For the tight-binder Lys-34 mutant CI + op, $k_\textrm{d}$ further decreases to $1.6\times 10^{-3}$ s$^{-1}$ by a factor of 3.

Apart from dissociation from locations along the length of the DNA, CI$_2$ may also dissociate at one of the ends of the DNA molecules. However, immobilization of one end via biotin-streptavidin link the the surface likely prevents dissociation from that end. Instead, reflection is possible, and the CI$_2$ returns to 1D diffusion. For nop DNA, we observed no length dependence of $k_\textrm{d}$. It is possible that the free DNA end is totally reflective, or 1D diffusion covers only a small range of DNA, so that CI$_2$ dissociates from a local sequence. For the first case, we would expect to see some $k_\textrm{d}$ length dependence for CI$_2$ and op DNA, since CI$_2$ on shorter DNA encounters more reflections. In the second case (1D diffusion covers a very short distance), there will be no length dependence for op DNA either, since CI$_2$ falls off the operator sequence before it can interact with the non-operator sequences. Our observation supports the second case. We also observed little length dependence of the tight-binder mutant Lys-34 CI$_2$ to op DNA.

Finally, as noted above, if $k_\textrm{a}$ involves 1D diffusion--and the evidence summarized here suggests it does not--then $L$-dependence should also appear in $k_\textrm{d}$ measured as a function of $L$, since for simple Poisson events (Fig.~\ref{fig:trace}C,D) the law of microscopic reversibility holds. That $k_\textrm{d}$ shows no such dependence for op sequences (Fig.~\ref{fig:comp}A,~\ref{fig:kd}) strengthens our view that 1D diffusion plays little if any role in target finding over a range of 78 to 500 bp.

Extensive 1D diffusion on non-specific DNA can slow the search to a rate that is even slower than 3D searching alone. However, in bacteria, target sequences for most repressors, including CI, lie very close to their coding sequences~\citep{Kolesov2007}. Moreover, the motion of single messenger RNA molecules in living cells was recently reported to follow a subdiffusion process~\citep{Golding2006}. This implies that the local concentration of repressor is higher -- and the search volume smaller -- than is often assumed. While $k_\textrm{a}$ is seen to increase with $L$ for binding to nop DNA (Fig.~\ref{fig:ka}A), we could not resolve any $L$ dependence for op DNA (Fig.~\ref{fig:ka}B and C).  This suggests to us that, although nonspecific binding of CI scaling with $L$ is readily seen, 1D diffusion is unimportant once an operator is present, which is more consistent with this scenario.

We note that proteins that are to bind and diffuse on DNA are often designed to clamp onto the DNA ~\citep{Lopez2006} as they slide along. For those transcription factors with multiple distant targets, they often have high molarities in the cell. In the case of CI, once it is bound to the O$_\textrm{R}$1 and O$_\textrm{R}$2, which lie adjacent to the \emph{cI} coding region, \emph{cI} expression is up-regulated, resulting in increasing concentrations of CI$_2$. The expression is not repressed until the low affinity binding site O$_\textrm{R}$3 is also bound~\citep{Meyer1975, Dodd2004}. Increasing concentration then increases the probability that CI$_2$ will binding to O$_\textrm{L}$ sequences 2.4 kbp away, as well as the operator sites on superinfecting $\lambda$ DNA. Recent results show that CI transcription is further regulated by DNA looping induced by CI$_2$ binding at O$_\textrm{L}$1, O$_\textrm{L}$2, O$_\textrm{R}$1 and O$_\textrm{R}$2~\citep{Anderson2008}. DNA looping pulls the far binding site O$_\textrm{L}$ also to the vicinity of \emph{cI} coding region, increasing the probability of CI$_2$ finding the two lower affinity sites O$_\textrm{R}$3 and O$_\textrm{L}$3.

In summary, we have used quantitative single molecule fluorescence microscopy to investigate the interaction of CI$_2$ with DNA molecules. We show that CI$_2$ can bind DNA at both operator and non-operator sequences. However, CI$_2$'s interaction with the operator sequence is not affected by the length of flanking non-operator sequences. Our data suggested that CI$_2$ finds its target operator sequence mainly by 3D diffusion.\\

\begin{small}
We thank R. H. Austin and X. Yang for fruitful discussions. This work was supported by National Institutes of Health Grant HG001506 (to E.C.C.) and by Science and Technology Centers Program of National Science Foundation No. ECS-9876771 (to Y.W. and E.C.C.). Y.W. also acknowledges support from Princeton University.
\end{small}

\bibliography{CIDNA_2}

\clearpage
\begin{table}[b]
   \caption{\label{primertable} Right primers used for each DNA sample}
\begin{doublespace}
\begin{tabular}{cc}\hline
DNA & Primer sequence\\ \hline
nop 78 bp	& 5'-CCGAATTCCTCGAGTCTAGAGG-3' \\
nop 100 bp	& 5'-TATCGATTTCGAACCCGGGG-3' \\
nop 200 bp	& 5'-GGAAACAGCTATGACCATGATTAC G-3'\\
nop 300 bp	& 5'-TAATGTGAGTTAGCTCACTCATTAGGC-3'\\
nop 400 bp	& 5'-CAATACGCAAACCGCCTCTC-3'\\
nop 500 bp	& 5'-GTGGATAACCGTATTACCGCCT-3'\\ 	
op 78 bp 	& 5'-CATGCGACGTCGGGC-3'\\
op 100 bp	& 5'-GAATTCCTCGAGTCTAGAGGAGC-3'\\
op 200 bp	& 5'-GCCAAGCTATTTAGGTGACACTATAGAA-3'\\
op 300 bp	& 5'-GGCACCCCAGGCTTTAC-3'\\
op 400 bp	& 5'-CGCGTTGGCCGATTCAT-3'\\
op 500 bp	& 5'-GAGTGAGCTGATACCGCTCG-5'\\ \hline
\end{tabular}
\end{doublespace}
\end{table}

\clearpage
\section*{Figure Legends}
\subsubsection*{Figure~\ref{fig:Intro}.}
Sample preparation and single molecule imaging of CI-DNA interaction.
(A) Schematic description of the PCR primers used here. The O$_\textrm{R}$1 sequence
was inserted into a vector. A common PCR left primer with a 5' biotin and an internal Alexa Fluor 488 was used in each PCR reaction. The right primers were selected so that different primers produce different DNA lengths. nop DNA samples were synthesized using the vector lacking the O$_\textrm{R}$1 sequence as the PCR template.
(B) Layout of 24 DNA patches in 4 channels defined by vacuum grease (yellow stripes) between two rectangular coverslips.  Each channel had a different [CI$_2$] (shown as different red opacities) and allowing us to compare different $L$ DNA at the same CI concentration.
(C) Schematic drawing of the reaction in one patch. Green fluorescently labeled DNA molecules were immobilized on the coverslip surface. In solution, and also bound on DNA were (tdimer:CI)$_2$ molecules (CI binds as a dimer). Both molecules were visualized by TIRF microscopy.
(D) Superposition of red and green fluorescence images. Green spots are individual DNA molecules fixed to the coverslip surface. Red spots are tdimer:CI molecules. Colocalized DNA and CI, the yellow spots, indicate CI-DNA binding.

\subsubsection*{Figure~\ref{fig:bleach}.}
Molecular bleaching.
(A) Average bleaching curve of all molecules in a field of view as a function of time. The Alexa Fluor 488 fluorescence decreased 10$\%$ and the tdimer fluorescence decreased 30$\%$ after 100 s of continuous illumination.
(B, C) Intensity-time traces of two typical Alexa Fluor 488 molecules. The intensity dropped to the background level abruptly.
(D -- G) Intensity-time traces of four typical tdimer:Cys-88 CI molecules. Compared with single Alexa Fluor 488 molecules, there was no abrupt large drop of intensity. With time, the intensities of these molecules decreased gradually.

\subsubsection*{Figure~\ref{fig:trace}.}
Dynamics of CI$_2$ interacting with DNA (WT CI and op 300 bp DNA are shown here).
(A) Time trace of red fluorescent signal $I$ vs. $t$ from a single op DNA molecule displaying the transitions $DP_0 \rightleftharpoons DP_1$.  Bright intensities correspond to binding of a single (tdimer:CI)$_2$. The bold line-segments are generated by the transition-recognition program that counts the events.
(B) A single DNA molecule could be bound by more than one CI$_2$. During 5 -- 9 min, this DNA was bound by two CI$_2$'s. We filter out multiple binding events and accumulate statistics of dark intervals $t_\textrm{dark}$ and the bright intervals $t_\textrm{bright}$ using traces as in (A).
The histograms of the distribution of $t_\textrm{dark}$ and $t_\textrm{bright}$ are shown in Panels C and D, respectively.  Fits to exponential distributions ${\cal P}(t_\textrm{dark})$ and ${\cal P}(t_\textrm{bright})$ (bold curves) assuming Poisson processes yields $\lambda_0 = [\textrm{CI}_2]k_\textrm{a}$ and $\mu_1 = k_\textrm{d}$. (Also see main text)

\subsubsection*{Figure~\ref{fig:hist}.}
Histograms of red fluorescence intensities from (tdimer:CI)$_2$ (Lys-34 CI in this figure) per green fluorescent spot (op DNA in this figure), indicating the number of CI$_2$'s per DNA. DNA lengths $L$ are labeled in each panel. These DNA molecules were exposed to the same [CI$_2$]. Two peaks are apparent in each panel. The narrow peak near zero intensity (aligned with the dashed line) is the DNA population with no bound CI$_2$. The second peak is the DNA population bound with one molecule of CI$_2$ ($n \geq 1$). As $L$ increases, the peak broadens and shifts to higher intensity, reflecting the increase in number of CI$_2$'s per DNA.
{A histogram of tdimer:Cys-88 CI intensities is plotted as a thick line in each panel as the dimer standard.}

\subsubsection*{Figure~\ref{fig:L34}.}
Evolution of the fluorescence intensities of Lys-34 CI bound to DNA versus DNA length $L$ and [CI$_2$]. Normalized frequencies of fluorescence in each intensity bin are shown as a heat map. The DNA is nop (op) in the upper (lower) panels. Each panel shows how the histograms change as $L$ steps through the values 78, 100, 200, 300, 400, 500 bp, with [CI$_2$] fixed at the value shown on the top. Blue represent zero frequency, while red represent high frequency. In each panel, as $L$ increases, the center of the second peak shifts to larger CI$_2$ number as the peak also gets broader, reflecting in increasing $\langle n\rangle$. Increasing [CI$_2$] at fixed $L$ also increases $\langle n\rangle$. At the same $L$ and [CI$_2$], op DNA captures more CI$_2$ than nop DNA.

\subsubsection*{Figure~\ref{fig:mean}.}
The mean number $\langle n \rangle$ of CI$_2$'s per DNA molecule plotted vs. $L$.
(A) WT CI and nop DNA, (B) WT CI and op DNA, (C) Lys-34 CI and nop DNA, (D) Lys-34 CI and op DNA. In each panel, $\langle n\rangle$ is calculated as the mean red fluorescent intensity per DNA site, normalized by the intensity of single CI$_2$. Straight lines are best linear fits to $\langle n\rangle$.  Within each panel, [CI$_2$] decreases $\sim$3-fold between data sets with successively smaller slopes.  The values (in nM) are 25, 11, 4.4, 1.7 (Panel A);  4.4, 1.7, 0.6 (B); 2.0, 0.71, 0.23, 0.07 (C); and 2.0, 0.71, 0.23, 0.07 (D).

\subsubsection*{Figure~\ref{fig:comp}.}
Direct comparisons of rate constants for binding of WT CI$_2$ to nop versus op DNA.  Panel A compares the dissociation rate constants $k_\textrm{d}$ for op DNA (solid symbols) with that for nop DNA (open symbols).  For each DNA population, 2 lengths were investigated in 3 runs at the same nominal [CI$_2$] of 11 nM (different symbols).  Panel B shows the association rate constants $k_\textrm{a}$ extracted from the same experiments.

\subsubsection*{Figure~\ref{fig:kd}.}
Comparison of the dissociation rate constants $k_\textrm{d}$ measured for binding of WT CI to nop DNA (solid symbols) and op DNA (open symbols), and binding of the Lys-34 CI to op DNA (asterisks).  Six DNA lengths were investigated (at fixed [CI$_2$]) in each case.  For WT CI + nop DNA, [CI$_2$] was 11 nM in both runs (solid triangles).  For WT CI + op DNA, [CI$_2$] was 11 nM in 2 runs (open squares and circles) and 1.7 nM in a third (open diamonds). In Lys-34 CI + op DNA, [CI$_2$] was 0.07 nM.  In each case, a grey band is centered at the mean value of $k_\textrm{d}$ while its half-width gives the variance.  The observed values of $k_\textrm{d}$ increase significantly as we go from Lys-34+op to WT+op, to WT+nop (successively weaker binding of CI$_2$). However, $k_\textrm{d}$ does not display a resolvable $L$ or [CI$_2$] dependence.

\subsubsection*{Figure~\ref{fig:ka}.}
The corresponding values of $k_\textrm{a}$ extracted from the series of experiments involving the CI-DNA populations A: WT CI + nop DNA. The two different symbols are for two different experimental runs. B: WT CI + op DNA, and C: Lys-34 CI + op DNA shown in Fig.~\ref{fig:kd}. Panel A shows that $k_\textrm{a}$ in nop DNA increases linearly with $L$, but in the 2 op DNA populations, $k_\textrm{a}$ is virtually $L$ independent. In Panel A, the nominal [CI$_2$] $\sim$ 11 nM for both open and solid triangles. In B, [CI$_2$] is 11 nM (open square), and 1.7 nM (solid circles).  In C, [(Lys-34 CI)$_2$] is 0.07 nM.

\begin{figure}
   \begin{center}
      \includegraphics*[width=7in]{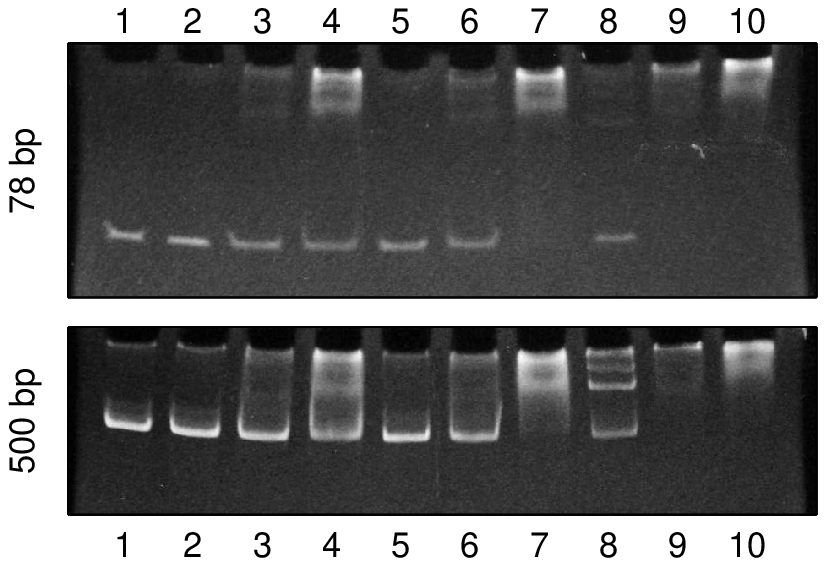}
      \caption{}
      \label{fig:gel}
   \end{center}
\end{figure}

\subsubsection*{Figure~\ref{fig:gel}.}
Gel-shift assays of DNA-CI reactions. Upper panel: 78 bp DNA. Lower panel, 500 bp DNA. DNA concentration in each lane were 20 nM. Lane 1: nop DNA control with no CI$_2$. Lane 2 to 4, nop DNA interacting with increasing concentrations of WT CI$_2$. Lane 5 to 7, op DNA interacting with increasing concentrations of WT CI$_2$. Lane 8 to 10, op DNA interacting with increasing concentrations of Lys-34 CI$_2$. The three CI$_2$ concentrations are 5 nM (Lane 2, 5, 8), 36.5 nM (Lane 3, 6, 9) and 452 nM (Lane 4, 7, 10).

\clearpage

\clearpage

\clearpage

\clearpage

\clearpage

\clearpage

\clearpage

\clearpage

\clearpage

\clearpage

\end{document}